\documentclass[conference]{IEEEtran}

\IEEEoverridecommandlockouts
\usepackage{cite}
\usepackage{amsmath,amssymb,amsfonts}
\usepackage{algorithmic}
\usepackage{graphicx}

\usepackage{textcomp}
\usepackage{xcolor}
\usepackage{tabu}

\usepackage{url}
\def\BibTeX{{\rm B\kern-.05em{\sc i\kern-.025em b}\kern-.08em
    T\kern-.1667em\lower.7ex\hbox{E}\kern-.125emX}}
    \makeatletter
\def\ps@headings{%
	\let\@oddhead\@empty
	\let\@evenhead\@empty
	\def\@oddfoot{\@IEEEheaderstyle\hfil\thepage}%
	\def\@evenfoot{\@IEEEheaderstyle\thepage\hfil\hbox{}}%
}

\makeatother
\begin{document}

\title{Exploring the Landscape of Fairness Interventions in Software Engineering\\}

\author{\IEEEauthorblockN{Sadia Afrin Mim}
\IEEEauthorblockA{\textit{Department of Computer Science} \\
\textit{George Mason University}\\
Fairfax, Virginia \\
safrinmi@gmu.edu}

}

\maketitle

\begin{abstract}
Current developments in AI made it broadly significant for reducing human labor and expenses across several essential domains, including healthcare and finance.  However, the application of AI in the actual world poses multiple risks and disadvantages due to potential risk factors in data (e.g., biased dataset).  Practitioners developed a number of fairness interventions for addressing these kinds of problems.  The paper acts as a survey, summarizing the various studies and approaches that have been developed to address fairness issues.
\end{abstract}


\section{Introduction}
With technological advancements, human civilization has witnessed various software innovations. Among these, the progress in artificial intelligence stands out. AI is now widely applied across essential sectors (e.g. healthcare, criminal justice, finance, and natural language processing)~\cite{iqbal2022assurance}. Artificial intelligence is widely employed in multiple domains to reduce human labor and improve efficiency by providing expedited outcomes~\cite{shah2019towards}.\\
AI models are trained on datasets collected from the real world. However, these datasets often contain biases against particular communities, (e.g., race, gender, religion)~\cite{ucl2024bias}. Training models on such biased data leads to unfair outcomes in AI applications. In addition to historical biases, there are other ways that AI models have been tainted.  Because AI is governed by a specific country, it may exhibit bias influenced by those policy makers~\cite{Peters2022}. Additionally, the dataset may contain fewer samples for a particular community. Additionally, the initial labeling of training data may be biased, as may the aggregation of models that were trained for a specific scenario and are now being used for generic purposes~\cite{haliburton2024investigating}. Numerous instances of AI bias have been reported in recent times, particularly in areas like loan applications~\cite{lipsitz2006possessive}, criminal justice~\cite{arowosegbe2023data}, image processing~\cite{udefi2023analysis}, and hiring processe~\cite{fitria2021gender}.\\
To address these challenges and ensure ethical outcomes in technological advancements, researchers have explored various approaches~\cite{lee2021landscape}. For instance, some have proposed fairness metrics to detect and measure bias~\cite{mehrabi2021survey}, while others have focused on bias mitigation strategies. Additionally, regulatory frameworks and ethical guidelines, such as the EU AI Act~\cite{edwards2021eu} and GDPR~\cite{li2019impact}, have been introduced to enforce responsible AI use. Researchers are also emphasizing transparent AI models, diverse and representative datasets, human-in-the-loop oversight, and explainability techniques to enhance trust and accountability in AI systems. Open-source communities have collaborated to integrate these efforts, developing fairness tools that assist practitioners in achieving fair AI outcomes.\\
Research communities have also explored the landscape of available fairness tools through human-centric interviews and qualitative studies. The article provides an extensive overview of the research conducted in this domain. We conducted a literature search using Google Scholar, the ACM Digital Library, and IEEE Xplore, employing keywords such as Fairness, AI Fairness, AI Bias, and AI Ethics. Our selection primarily includes literature from 2018 to 2025, as the concept of ethical AI development has gained significant attention during this period. The collected studies come from top-tier conferences, including the International Conference on Software Engineering (ICSE), the CHI Conference on Human Factors in Computing Systems, and the ACM Conference on Fairness, Accountability, and Transparency (FAccT). Our list also includes Journals (e.g.,ACM computing surveys, Proceedings of the ACM on Human-Computer Interaction, Towards Trustworthy Artificial Intelligent Systems(Springer)). As AI moving forward, communities are continually engaging in this field to address potential risks and challenges, ensuring safer and more responsible AI outcomes.
The main contributions of this work are as follows:
\begin{itemize}
    \item Cataloging existing research studies on Ethical AI tools.
    \item Providing insights from existing studies to guide future research directions.
\end{itemize}
The literature is organized as follows: the section~\ref{preli} discusses necessary preliminary concepts to understand open source AI ethics tool development. The taxonomy of the ethical AI tool research area is covered in section~\ref{area}, and the taxonomy from the earlier section is covered in section~\ref{taxon} to structure my understanding of the work that has been done in this field. The section~\ref{open} discusses some open research problems in this area according to my understanding \& section~\ref{conclusion} concludes the survey.
\section{Preliminary}
\label{preli}
In this section, I explore the key factors essential for a deeper understanding of this survey, focusing primarily on ethical software development practices \& its related aspects.
\subsection{Fairness \& Ethics in Software}
Fairness and ethics play a vital role in our daily lives. As technology advances, the need to ensure ethical outcomes from these developments has become increasingly important. With AI evolving, creating new models, and becoming more integrated into our lives, various interventions have been introduced to support practitioners in this field. The concepts of fairness and bias are influenced by several key factors, which I will discuss in detail in the following subsections.
\subsubsection{Protected Attributes}
Specific characteristics that are ethically or legally shielded against unfair treatment in AI systems are known as protected attributes~\cite{chen2024fairness}. Typically, these attributes are associated with societal bias or historical injustice~\cite{zajko2021conservative}. Most commonly considered protected attributes include factors(e.g., race, sex, age, disability status, religion, socio-economic status)~\cite{chen2024fairness}. Protected attributes are divided into privileged and unprivileged categories according to their description.  Because of historical biases in datasets, privileged groups usually gain from AI predictions, while underprivileged groups struggle.  For example in Amazon recruiting software scenario  where sex is the protected attribute. Male applicants comprised the privileged group and female applicants the unprivileged group. This resulted in a bias in hiring that privileged male candidates~\cite{fitria2021gender}. Since many laws have been put in place over time to encourage equity and avoid prejudice in a variety of fields (e.g., Fair Housing Act~\cite{schwemm1978discriminatory}, Civil Rights Act~\cite{act1964civil}), it is essential that these AI applications be fair.
\subsubsection{Sources of Bias}
Bias in systems, particularly AI, manifests as unfair favoring or discrimination against specific groups, often based on race or gender, stemming from various sources. Historical bias, where existing real-world data is inherently skewed, can be built in the system. Data labeling bias occurs when training data is contextually limited, leading to poor performance in other scenarios. Sampling bias arises from using underrepresentative datasets, while automation bias can amplify existing prejudices through over-reliance and repeated retraining on biased data. Furthermore, Moreover, authoritative bias may arise when software governed by one group in particular displays partiality towards competing groups.
\subsubsection{Fairness Metrics}
Fairness metrices are used to detect \& measure bias in models based on some mathematical methods~\cite{garg2020fairness}. Researchers have proposed a variety of fairness metrics that can be used with AI models in different scenarios.  AI Fairness 360, for instance, offers over 70 fairness metrics~\cite{bellamy2019ai}. Here is an example of mathematical notation of fairness metrices \textit{Equal Opportunity} where TP denotes true positive and FN signifies false negative. For two groups, the true positive rates (TPR) should be equal~\cite{pagano2023bias}.

\[
\frac{\text{TP}_1}{\text{TP}_1 + \text{FN}_1} = \frac{\text{TP}_2}{\text{TP}_2 + \text{FN}_2}
\]

Among the available fairness metrics most commonly used ones are: Demographic Parity, Equalized Odds, Equal Opportunity, Predictive Parity, Disparate Impact, Calibration~\cite{garg2020fairness}. These metrics can be employed in a variety of contexts. For instance, we can use Demographic Parity for group fairness, while practitioners use Fainress through Awareness~\cite{castelnovo2022clarification} for individual fairness.

\subsubsection{Bias Mitigation Algorithms}
Fairness research is not limited to identifying and quantifying bias; it also focuses on reducing potential bias. Many bias mitigation strategies have been created by researchers to assist in the fair implementation of AI models~\cite{wang2020mitigating}. According to their adoptation during various phases of ML lifecycle they have been classified into three categories~\cite{chen2023comprehensive}.
\begin{itemize}
    \item \textbf{Preprocessing:} Preprocessing bias mitigation approaches are applied before training a model and primarily focus on redistributing and adjusting data. One such technique is \textit{Disparate Impact Removal}, which modifies data distribution while preserving the ranking within groups~\cite{feldman2015certifying}.
     \item \textbf{Inprocessing:} Inprocessing bias mitigation approaches are applied during training a model and focus mainly on adjusting algorithms by tuning hyperparameters. One example of inprocess bias mitigation algorithm is \textit{Prejudice Remover}, where they add extra parameter and tune the model to obtain fair outcome~\cite{kamishima2012fairness}.
      \item \textbf{Postprocessing:} Postprocessing bias mitigation approaches are applied after training a model and focus mainly on adjusting the prediction rather than retraining the model~\cite{lohia2019bias}. One example of postprocess bias mitigation algorithm is \textit{Reject Option Classification}, where they adjust the probability threshold to obtain fair outcome~\cite{kamiran2012decision}.
\end{itemize}
\subsection{Open Source Fairness Tools}
The goal of ethical software development is to provide technology that complies with the values of responsibility, transparency, fairness, and user privacy.  To safeguard people and stop exploitation, developers must take user permission, data security, and regulatory compliance into account.  While sustainability promotes ecologically conscious development methods, inclusivity and accessibility guarantee that software is used by a variety of demographics.  Avoiding algorithmic bias, maintaining responsible data handling, and guaranteeing open-source integrity are other components of ethical decision-making.  Frequent audits, human supervision, and ethical standards compliance all aid in risk mitigation and guarantee that software solutions minimize harm while fostering beneficial social influence.

As fairness became a growing concern with technological advancement, open source communities have worked to build various software interventions (e.g., tools, frameworks) to ensure easier access to those fairness artifacts. In this study, we specifically focus on interventions that guarantee reliable and trustworthy outcomes from AI model predictions. Our previous work~\cite{mim2023taxonomy} have shown a taxonomy of 14 existing fairness toolkits. We collected the keywords pertaining to ethics \& fairness used in those repositories from the \textbf{About} section of their GitHub repository \& further discovered 48 additional fairness repositories that are explored further in the work. We also analyzed their status as active/inactive in open source based on last activity in open source (e.g., issue resolution, commit, pull request) following methodology from prior work~\cite{coelho2017modern}. Only 20 out of 62 fairness repositories are actively maintained from our list.

It is evident that open-source communities have actively contributed to developing fairness interventions (e.g., toolkits, libraries) to support practitioners in building trustworthy AI models. Newer models have emerged, and community efforts have focused on aligning these models with fairness definitions. Consequently, in our curated list of 62 fairness repositories, we identified works addressing a diverse range of models, from traditional classification algorithms to advanced LLMs. The development and adoption of these models must be fair and equitable for all. Thus, open-source initiatives continue to strive to integrate emerging models and technologies to promote equity and fairness in software systems.

Upon analyzing the current literature, we identified tools that offer assistance for both conventional models (e.g., Random Forest, SVM, Logistic Regression) and modern architectures (e.g., GPT, LLMs, Adversarial Networks).
They showed various distinguishing factors which are listed here.
\begin{itemize}
    \item \textbf{Detection vs Mitigation:} 
    Certain tools employ fairness metrics to evaluate the possible existence of bias in a model, while others also incorporate mitigation techniques using various fairness algorithms discussed earlier in this section. The majority of interventions offer both bias detection and mitigation support (e.g., Microsoft Fairlearn); however, we identified some tools that focus solely on general mitigation without detecting bias. Additionally, we found a significant number of open-source fairness tools dedicated exclusively to detecting and measuring bias (e.g.,Microsoft SafeNLP).
    \item \textbf{Guidance vs Analysis:} These open-source tools fall into two categories based on their functionality: Analysis tools and Guidance tools~\cite{mim2023taxonomy}.  The main goals of the guidance tools are to develop ethical checklists, define fairness measures, and establish parameters to measure fairness(e.g., Drivendataorg Deon).  On the other hand, analytical tools use predetermined metric values to assess and measure the fairness of a model (e.g., Dssg Aequitas).
\end{itemize}
\section{Area Taxonomy}
\label{area}
The taxonomy section of the survey investigates the landscape of fairness interventions. By analyzing previous research on ethical software development specifically in  AI-driven areas, the taxonomy was developed. Fairness research is progressing over time. Researchers have investigated to make technologies robust \& ethical for real-world practices. 
We discovered several initiatives to comprehend and resolve biases in software results during our investigation of ethics and fairness in software systems. Numerous software tools have been created in various fields to examine these biases and suggest quantifiable methods for reducing them. Researchers are focused on how bias enters software systems, from data collection to algorithm design. Assessing AI models increasingly emphasizes fairness and ethical considerations. To explore ethical issues in AI software, researchers used industry surveys, data management analyses, interventions and human-centered studies. To identify and reduce bias in machine learning models, for example, IBM's AI Fairness 360 toolbox provides a wide range of indicators and methods. Interestingly, some of these systems let users create their own ethics checklists, allowing them to customize the study to address certain ethical issues.  Researchers have also investigated how these interventions have changed throughout time, offering insightful information to guide the creation of workable solutions. For example, the Deon, an open-source command line interface, enables users to their own ethics checklist in a project.
 These developments demonstrate the continuous efforts to design more transparent and equitable AI systems. We examined existing research on developing ethical and trustworthy systems, while also aiming to understand the efforts of open-source communities in this area. My understanding of literature study in this are can be classified as follows:
 \begin{itemize}
 \item Ethical \& fairness concerns for generic software development
    
     \item Fairness concerns for AI-assisted software systems 
     \item Human-centric evaluation of fairness in software systems
     
 \end{itemize}
 
\section{Taxonomy-based survey}
\label{taxon}
Researchers have explored the domain of ethical AI and fairness to ensure their effective and reliable integration into applications.  Given that many modern software applications are AI-driven, this study investigates various approaches categorized in the preceding subsection, focusing on fairness in AI-based software. This section presents a comprehensive analysis of the reviewed literature in this field and the summary of the related literature to relevant work is shown in Table~\ref{summary}.
\subsection{Literature on Ethical Software Engineering}
To attain equitable and ethical results, researchers have introduced a variety of interventions that address ethical concerns. In the context of the software development lifecycle, ensuring ethics and fairness is relevant across multiple phases. During the planning and design phases, it is essential to establish requirements and objectives that promote fairness.\\
\subsubsection{Practices of DEI in software engineering}
In addition, ethical concerns should be addressed during the verification and validation phases of software development. 
Daniela et al. in their book aims to promote equity, diversity, and inclusion (ED\&I) in software engineering by analyzing challenges, initiatives, and best practices to ensure ethical software development~\cite{damian2024equity}. The book is divided into 6 parts where they discussed various aspects of ethical software engineering. They discussed general problems and obstacles minority groups experience in the IT industry, as well as how software can help them overcome these obstacles.  They also addressed inclusivity in software products (such as the inclusion of non-binary individuals in systems). Later, they focused on diversity in development teams (e.g.,gender diversity, intersectionality).  They introduced the required measures, such as rules and interventions, that have been adopted to achieve equity.  The interventions for computer science and software engineering education are also emphasized. These include mentorship programs for women in open source, gender-inclusive hackathons, and the role of codes of conduct in open-source communities.  Moreover, insights about research approaches for examining diversity in software engineering are presented in their discussion.  It covers methods for measuring diversity, gathering and evaluating demographic data in an ethical manner, and promoting underrepresented perspectives in studies.\\
\subsubsection{Ignorance and prejudice in software engineering}
Fairness and diversity in software engineering research have been the subject of numerous studies.  Yuriy et al. provided a vision for software engineering research's role in promoting justice while highlighting the importance of this field in addressing ethical issues~\cite{brun2018software}.  The main obstacles to creating just and equitable software are addressed in this paper by outlining specific steps that software engineering researchers might take.  They cited participation in projects like the IEEE/ACM Workshop on Software Fairness as evidence of the software engineering community's increasing ethical consciousness (FairWare 2018).  They also found that interdisciplinary cooperation, especially with the machine learning community, is a driving force behind the advancement of software system fairness.  As crucial research areas, the report also suggested defining fairness targets, automating testing procedures, and creating and implementing fair test suites as crucial components of ethical software engineering. \\

\subsubsection{Interventions for fairness in software}
Researchers have also highlighted the importance of data and data-driven software in guaranteeing equitable systems, in addition to the components of an ideal software development technique.  In their study, Johnson et al. looked at the moral dilemmas raised by data-driven software and offered solutions~\cite{johnson2021towards}.  Their research examined previous interventions and literature, covering both theoretical models and practical applications. They noted a number of important areas that needed work.  First, they raised concerns about practitioners' limited knowledge of ethical issues in this area, pointing out that although there is research on ethical software development, it is still lacking.  They also emphasized how crucial it is for software teams to have both external and internal assistance to create increasingly sophisticated and morally sound systems. Furthermore, other stages of data-driven software development also need attention, even if the majority of data-driven workflows are in line with machine learning pipelines.  The study emphasized the necessity of putting current ethical standards into practice in actual systems as opposed to only talking about them in theory.  Furthermore, fairness in computers has frequently been the focus of ethical issues; however, to guarantee a whole approach, other crucial ethical aspects must be examined.  Researchers should evaluate a project's ethical consequences holistically rather than just using accepted procedures. Additionally, they underlined the necessity of interventions that go beyond theory to address practical needs.  It is imperative that current ethical treatments be used more widely and practically.  The study offered a thorough summary of current status of ethical research in data-driven software development by identifying these open research problems. 

Zhang et al. looked at the problems of bias and ignorance in the development of data-driven software, especially machine learning-based initiatives in terms dataset \& feature size~\cite{zhang2021ignorance}.  Their study evaluated how feature set and dataset sizes affected software performance fairness.  Their results show that adding more characteristics can increase system fairness by up to 38\% and that a larger dataset can improve fairness within a project.  They did, however, also draw attention to an important risk, a system with a notably small feature set may act in more unpredictable and sometimes dangerous ways, even with a huge dataset.

To create reliable software systems, researchers have also looked into a variety of interventions and solutions.  The PICSE framework for building software tool trust was put forth by Johnson et al. in their study~\cite{johnson2023make}. This framework aligns with the requirement analysis, designing \& verification phases of software development. To find out their thoughts on creating dependable and trustworthy software, they interviewed eighteen professionals in the field.  They created a framework that incorporates important elements that improve system reliability based on these discoveries. The framework highlights a number of important components.  Collaboration between internal and external parties is crucial in ensuring that software development is in line with validation procedures that improve user acceptance.  Interactions are another crucial component since a system builds trust by integrating user experiences and insights.  Furthermore, a key factor in controlling ethical aspects of the system is control, which is decided by its developers.  The paradigm also recognizes that defining system trustworthiness requires the use of cutting-edge assessment techniques, such as determining correctness and consistency.  Furthermore, building confidence requires that system development be in line with engineers' expectations. All things considered, the PICSE framework emphasizes the key factors that practitioners need to take into account while creating a reliable software system. 

Phases of software testing and verification also require more attention.  Themis, a test suite generator created by Angell et al., is intended to identify prejudice and discrimination in software~\cite{angell2018themis}.  They created this technique to find any possible prejudice that a system might display because software testing is a vital step in the development process. Previously, the authors created a methodology known as causal discrimination that evaluates the connection between program behavior and sensitive properties~\cite{galhotra2017fairness}.  They also incorporated automated test suite development and bias reduction techniques.  They assessed and improved system equity by evaluating 20 real-world software systems using fairness testing approaches after establishing fairness testing criteria and introducing Themis.
\subsection{Literature on Ensuring Ethical Output from AI Models:}
AI is now integrated into various applications, including software engineering, making it crucial to ensure ethical outcomes from AI models.\\ 
\subsubsection{Fairness Practices in AI-driven Areas}
The study by Maria et al. investigated the complex nature of reliable AI, looking at methods to guarantee dependability using certification and benchmarking tools and suggesting frameworks for real-world AI use~\cite{franse2022practical}.  They made the case for AI development and governance that puts human values first by highlighting the serious problem of bias in domains including jobs, healthcare, and social contexts.
By establishing a taxonomy of fairness elements based on actual instances of biased AI systems, Mehrabi et al. investigated fairness in AI~\cite{mehrabi2021survey}.  They discussed ways to assess fairness,  (e.g., equal opportunity and demographic parity) and looked at a variety of bias sources, including biases in data and algorithms.  They also examined a number of bias mitigation techniques intended to advance equity in AI.  Their research also signified important factors to maintaining equity, such managing complex scenarios and defining precise fairness standards. 
Pessach et al. emphasizes the necessity of objective AI, where they noted that ML can magnify societal prejudices based on historical data, influencing important judgments in fields like healthcare and justice~\cite{pessach2022review}. Sources of unfairness in data, algorithm design, and decision-making are identified by the review.  They explored the internal structure of several terminologies of fairness, such as fairness metrices. After assessing each method's advantages and disadvantages, the authors divide fairness action phases in three categories: pre-processing, in-processing, and post-processing.  To demonstrate the wide range of the problem, the study also discusses new fairness research in  diverse sectors (e.g., computer vision, word embeddings). 
With an emphasis on bias, assessment, and correction, Caton and Haas's work provides a comprehensive understanding of fairness in machine learning~\cite{10.1145/3616865}.  They drew attention to the fact that ML frequently exhibits gender, racial, and disability prejudices when it is applied in crucial judgments.  The poll divides fairness solutions into three categories: pre-training data change, training algorithm modifications, and post-training prediction corrections.  Eleven fairness strategies, including reweighing and causal procedures, are categorized by the authors. They talk about the constraints of fairness measurements like equalized odds and demographic parity. The survey also examines existing fairness toolkits and discusses fairness in domains other than binary classification, such as regression and recommendation systems. The study emphasizes major challenges in fairness research, including the trade-offs between fairness and accuracy, along actual utilization of fairness.

By examining data, evaluation strategies, corrective procedures, and software tools, Pagano et al. looked over bias in ML models. 45 relevant research articles from 128 published between 2017 and 2022 were examined after an extensive review~\cite{pagano2023bias}. This research's main goal is to detect and mitigate bias in datasets from fields like criminal justice and healthcare by applying statistical techniques and fairness metrics.  Usually, data, algorithm modifications, or result post-processing are used to mitigate bias resulting from data, algorithms, and user interactions.  They also mentioned about Aequitas being a widely used tool among the fairness tools even after availability of many fairness tools. The emphasis on binary classification, the disregard for multiclass issues, and the inconsistent use of fairness criteria in various applications are some of the main drawbacks of the current study.  This demonstrates the necessity for additional study on model design to find suitable fairness criteria to particular use cases. The assessment also highlights how crucial it is becoming to make algorithms understandable and transparent to maintain equity.
\subsubsection{Comparison of Existing Interventions}

The effectiveness of current AI ethics tools in putting ethical ideas into practice is assessed in Ayling et al~\cite{ayling2022putting}. By comparing them with well-established practices from other fields, such as environmental impact and financial audits, they analyzed tools designed to address important issues like bias and transparency. According to their research, actual transparency is hindered by the fact that the majority of AI ethical tools are designed for internal rather than external verification.  These technologies frequently lack transparency and are ineffective at mitigating standardized prejudice.   They frequently leave out important parties, including impacted consumers, which reduces their efficacy.  The authors suggest that accountability and participation techniques from other domains must be incorporated into AI impact assessments. To guarantee that AI ethical tools are beneficial, the article highlights the need for proper governance and regulatory supervision.  It promotes external responsibility, increased stakeholder involvement, and matching AI ethical tools with validated impact assessment and auditing frameworks.
Researchers also investigated the landscape of interventions developed in this area.Nguyen et al. assessed 41 tools with an emphasizing on practical insights of documentation, and ease of use~\cite{nguyen2024literature}. Many tools have flaws according to their analysis, including poor user interfaces, insufficient documentation, and inadequate flexibility for various data types.  Additionally, a lot of tools are only made for particular datasets and don't receive regular updates, which limits their generalizability. Although several tools have benefits like affordability and system compatibility, the majority need significant enhancements to satisfy industry standards.  The study emphasizes the critical need for more flexible, well-documented, and user-friendly fairness assessment instruments.  To ensure fairness in AI-driven software, these advancements are essential for empowering developers to recognize and mitigate prejudice.

Wong et al. analyze AI ethics toolkits and their underlying requirement regarding moral AI behavior~\cite{wong2023seeing}. To learn how they define ethical concerns and what they anticipate implement ethical work examined 27 toolkits.  They discovered a significant gap between the limited availability of these tools and the challenges and ethical problems encountered by AI practitioners. The majority of toolkits ignore broader social and organizational factors.  They frequently do not offer helpful advice on involving stakeholders, resolving power disparities, and encouraging group action.  They see ethical problems as just technical problems. According to the authors, these toolkits usually serve to uphold current power dynamics, hindering actual ethical advancement.

Mim et al. developed a taxonomy of 14 existing fairness interventions, categorizing them based on their features and usage within the machine learning workflow~\cite{mim2023taxonomy}. They organized these interventions into three stages—preprocessing, in-processing, and post-processing—to assist practitioners in identifying the most suitable techniques for each phase of model development. This study examines the degree to which fairness tools from research works are applicable in the software sector. 
\subsubsection{Interventions for ML fairness in Software}
The usage of IBM AI Fairness 360, a toolkit that comprises 70 fairness metrices and 10 bias mitigation techniques, is highlighted in a large portion of the literature among the technologies under analysis~\cite{bellamy2019ai}.  For assessing fairness, researchers and practitioners  make substantial use of AI Fairness 360.  Practitioners are encouraged to contribute and expand the toolkit's capabilities by its robust open-source community, which actively supports and maintains it.  Because of its widespread use and ongoing maintenance, it is regarded as a reliable source for evaluating the fairness of AI systems.  On top of AIF360, further interventions have been created.  For example, Johnson et al. built a toolkit called Fairkit-learn on top of Scikit-learn and AI Fairness 360~\cite{johnson2020fairkit}.  They added more metrics and automatic search to the fairness features.
\subsection{Literature on human-centric study on ethical AI}
Given that preliminary surveys and studies have emphasized the importance of human validation in understanding the broader ethical implications of technology, researchers have explored the regulatory landscape and interventions for fairness through various human-centric studies.
Lee and colleagues interviewed industry practitioners to assess the strengths and weaknesses of six widely recognized fairness tools~\cite{lee2021landscape}. Their study compared these tools based on their open-source attributes, including programming language, licensing, and implemented algorithms. Through these interviews, they found that these tools have a steep learning curve, making them challenging to adopt. Additionally, they observed that fairness considerations are minimally addressed throughout the end-to-end model development lifecycle and that these tools are not well-integrated into current workflows, limiting their practical implementation.

Interviews conducted by Widder et al. to examine the ethical reasoning of participants in an open-source, AI-powered Deepfake project~\cite{widder2022limits}.  According to their findings, deepfakes are frequently misused to produce damaging content, such as false information and non-consensual videos.  Even while the open-source community is aware of these problems, strict regulations can lead to uncontrolled project duplications, which hinders their attempts to govern them.  They also highlighted the gap between open-source AI's benefits.  Developers struggle to balance the fundamental principles of open-source development with the requirement for ethical actions.

In a study involving interviews with 20 machine learning practitioners~\cite{richardson2021towards}, Richardson et al. investigated the practical application of fairness toolkits such as TensorFlow Fairness Indicators~\cite{richardson2021framework} and Aequitas~\cite{saleiro2018aequitas}. Utilizing a self-developed rubric, they evaluated these tools based on interview feedback, revealing two primary areas for improvement. Firstly, the study underscored the necessity for fairness analysis tools to be more inclusive, supporting a wider variety of machine learning models. Secondly, the researchers emphasized the critical need to enhance the usability of these toolkits to facilitate easier adoption and integration into practitioners' workflows.

Deng et al. in their work analyzed how industry practitioners integrate fairness toolkits in their workflow interviewing 11 ML practitioners~\cite{deng2022exploring}. Their analysis revealed the gap between the need of fairness practitioners \& also revealed the challenge of beginners using these tools. The study participants expressed a need for improved educational resources integrated within fairness toolkits to enhance their understanding of fairness concepts. To effectively address fairness in machine learning, they showed the need to foster interdisciplinary collaboration, e.g., legal, ethical, and business perspectives.

Madaio et al. conducted a study involving interviews with 33 AI practitioners across 10 teams from three organizations to understand how they evaluate fairness in machine learning systems~\cite{madaio2022assessing}. Their research uncovered the processes, challenges, and support needs of these practitioners. Key findings included difficulties in selecting appropriate fairness metrics, a lack of sufficient stakeholder engagement, and challenges in the data curation process. The authors further emphasized the need for improved frameworks, organizational support, and resources to address these issues and facilitate better fairness assessments.
Researchers investigated human-centered AI guidelines to make AI products more sophisticated. Yildirim et al., through interviews with 31 industry professionals spanning various roles, identified a significant gap: current AI guidelines fail to adequately support the initial design phases of AI systems~\cite{yildirim2023investigating}. Their research revealed that practitioners struggle to locate and apply relevant guidelines during early project development. Consequently, the authors emphasized the need for domain-specific, adaptable human-AI guidelines to enhance the development of trustworthy AI products.

\begin{table*}[ht]
\caption{Overview of Related Literature \& Corresponding Research Areas }
\centering
\begin{tabular}{| l | c |}
\hline
\textbf{Workflow Relevance} & \textbf{Related Literature} \\
\hline

\multicolumn{2}{|c|}{\textbf{Ethical Software Development}} \\
\hline
Practices of DEI  in software engineering   & \cite{damian2024equity} ,~\cite{brun2018software}, ~\cite{johnson2021towards}\\
\hline
Ignorance and prejudice in software engineering & \cite{zhang2021ignorance}\\
\hline
Interventions for fairness in software & \cite{johnson2023make}, \cite{angell2018themis}, \cite{galhotra2017fairness}\\
\hline
\multicolumn{2}{|c|}{\textbf{Fairness in AI-Driven Software Engineering.}} \\
\hline
Fairness practices in AI-driven areas   & \cite{franse2022practical}, \cite{mehrabi2021survey}, \cite{pessach2022review}, \cite{10.1145/3616865}, \cite{pagano2023bias} \\
\hline
Comparison of existing interventions & \cite{ayling2022putting}, \cite{nguyen2024literature}, \cite{wong2023seeing}, \cite{mim2023taxonomy}\\
\hline
Interventions for ML fairness in software & \cite{bellamy2019ai}, \cite{johnson2020fairkit}, \cite{saleiro2018aequitas}\\
\hline
\multicolumn{2}{|c|}{\textbf{Human-centric Studies for Fairness}} \\
\hline
Studies on the implementation of fairness interventions & \cite{lee2021landscape}, \cite{richardson2021towards}, \cite{deng2022exploring}, ~\cite{madaio2022assessing}\\
\hline
Studies on the ethical guidelines of AI & \cite{yildirim2023investigating},\cite{widder2022limits}\\
\hline
\end{tabular}
\label{summary}
\end{table*}

\section{Open Problem}
\label{open}
In the previous sections, we reviewed literature based on our proposed taxonomy and examined how researchers have explored various dimensions of fairness and ethical considerations in software development. This section highlights open research challenges in this domain, which could inform potential future research directions.
\subsection{Software Engineering Open Research Direction}
The current landscape of ethical research in software engineering can be expanded further to enhance reliability. Here are some key insights.
\subsubsection{Broadening Ethical Concerns in Software Engineering}
Although ethical software engineering has advanced, a consistent definition of 'fairness' is lacking. To address this, future research should focus on developing standardized ethical frameworks. These frameworks must be broadly applicable, incorporating the diverse needs of various domains, stakeholders, and software systems, thereby fostering greater inclusivity in software development. Additionally, the majority of the studies have focused on contemporary approaches and interventions. Long-term implications are therefore required in research. For instance, autonomous vehicles are becoming increasingly popular these days~\cite{martinho2021ethical}. Very little research in this field has focused on ethical issues. 
\subsubsection{Disconnect between Theoretical Concepts with Real-world Applications}
Johnson et al. highlighted a significant gap between ethical considerations in theory and their practical implementation~\cite{johnson2021towards}. Research should prioritize the development of actionable ethical guidelines and tools for practitioners. Although ethical discussions and research are increasingly prevalent in academic settings, there have been limited efforts to translate them into real-world applications. For instance, there is a need for robust frameworks to ensure ethical practices in data-driven technologies. Additionally, ethical concerns within software engineering pipelines should be addressed through more aligned solutions that consider key factors such as resources and algorithms.
\subsection{AI-based Software Engineering Research Directions}
With industries rapidly adopting AI, several open research challenges in AI-driven software engineering and tools are explored here.
\subsubsection{Expanding Fairness beyond Traditional Models}
Open-source communities have developed numerous fairness tools over time, incorporating various metrics and algorithms~\cite{wong2023seeing}. Our previous analysis revealed that most interventions primarily target traditional classification models, such as binary classification. However, many fairness interventions are also applicable to multiclass classification. As the industry evolves and adopts more complex algorithms, fairness interventions must become more inclusive of these advanced models(e.g., GPT models). Additionally, greater emphasis should be placed on preprocessing bias mitigation in existing technologies, as research suggests that ethical considerations in AI should be addressed early in the development phase~\cite{yildirim2023investigating}. Therefore, further research is needed to expand this area.
\subsubsection{Developing Comprehensive Fairnes Tools}
Research has shown that existing fairness tools have a steep learning curve, making it difficult for beginners to effectively utilize them~\cite{lee2021landscape}. Consequently, further in-depth research is necessary to develop better support guidance tools for practitioners.
Additionally, these tools often lack the flexibility needed for easier integration into user workflows. The fairness metrics they provide are sometimes insufficient and may not always align with the specific criteria and objectives of a given project~\cite{gao2024fair}. Therefore, further research is needed to refine these criteria and ensure they are properly established to meet user needs and align with the machine learning lifecycle.
\subsubsection{Maintenance \& Open Source Engagement Analysis:} Although significant work has been done on fairness definitions, evaluation, and intervention development, there remains a notable gap in analyzing the engagement and maintenance of fairness tools in open-source communities. Researchers have employed various methodologies to assess the status of open-source repositories. For instance, Coelho et al. explored classification approaches for repositories~\cite{coelho2017modern}. To evaluate the health of a repository, it is essential to analyze open-source activity patterns (e.g., forks, pull requests, issue resolution) to understand how fairness tools are being adopted, how long they remain active, and the extent to which the open-source community integrates fairness into their work.
\subsection{Expanding Human-Centric Research on Fairness}
Our taxonomy highlighted a significant body of work on fairness aspects through human-centric studies. However, we identified several notable gaps in this area. Further research is needed to develop human-centered approaches to fairness that actively engage stakeholders from diverse backgrounds in the design and evaluation of AI systems. One key gap lies in the selection of interview participants.Most studies have focused on individuals from specific industries, and evaluations have primarily been conducted on widely recognized, industry-developed fairness interventions. Additionally, many studies have assigned participants specific tasks rather than exploring their personal experiences with these tools. Research indicates that users often prefer to share their personal experiences with a tool rather than simply completing predefined tasks~\cite{nguyen2024literature}.  

To address these gaps, more human-centered research is required to capture users' actual experiences, including insights on lesser-known fairness tools that have not been incorporated into previous study designs. Furthermore, research in this domain can be made more rigorous through stronger collaboration between academia and industry, as seen in projects like TensorFlow~\cite{abadi2016tensorflow}. However, we found a lack of human-centric studies led by academic researchers in this direction. Therefore, expanding research on industry-academia collaboration is crucial to fostering the development and broader adoption of fairness tools.
\section{Conclusion}
\label{conclusion}

This research aims to advance ethical software development, particularly within AI-driven applications. Through an analysis of existing work, I highlighted open research problems that can serve as a foundation for broader investigations in this area.
\nocite{*}

\bibliographystyle{plain} 
\bibliography{bibliography}

\end{document}